\documentclass[preprint]{elsarticle}
\usepackage{amsmath}
\usepackage{epsfig}
\usepackage{amssymb}
\usepackage{graphicx}%

\newcommand{\be}{\begin{equation}}
\newcommand{\ee}{\end{equation}}
\newcommand{\ba}{\begin{eqnarray}}
\newcommand{\ea}{\end{eqnarray}}

\begin{document}

\title{\textbf{One-dimensional Cooper pairing}}
\author[if]{R. Mendoza}
\ead{ramen@fisica.unam.mx}
\author[if]{M. Fortes\corref{cor1}}
\ead{fortes@fisica.unam.mx}
\author[im]{M. de Llano}
\ead{dellano@servidor.unam.mx}
\author[if]{M.A. Sol\'{i}s}
\ead{masolis@fisica.unam.mx}

\address[if]{Instituto de F\'{\i}sica, UNAM, Apdo. Postal
20-364, 01000 M\'{e}xico D.F., MEXICO}
\address[im]{Instituto de Investigaciones en Materiales, UNAM, Apdo. Postal 70-360, 04510 M\'{e}xico D.F., MEXICO}
\cortext[cor1]{Corresponding author}

\begin{abstract}
We study electron pairing in a one-dimensional (1D) fermion gas at zero
temperature under zero- and finite-range, attractive, two-body interactions. The binding energy
of Cooper pairs (CPs) with zero total or center-of-mass momentum (CMM)
increases with attraction strength and decreases with interaction range for
fixed strength. The excitation energy of 1D CPs with nonzero CMM display
novel, unique properties. It satisfies a dispersion relation with \textit{two}
branches: a\ phonon-like \textit{linear }excitation for small CP CMM; this
is followed by roton-like \textit{quadratic} excitation minimum for CMM
greater than twice the Fermi wavenumber, but only above a minimum threshold
attraction strength. The expected quadratic-in-CMM dispersion \textit{in
vacuo }when  the Fermi wavenumber is set to zero is recovered for \textit{any%
} coupling. This paper completes a three-part exploration initiated in 2D
and continued in 3D.
\end{abstract}
\maketitle
\section{Introduction}
Now just over a half-century old, the 1957 Bardeen, Cooper and Schrieffer
(BCS) theory of superconductivity \cite{bcs} is rightly regarded as one of
the most striking achievements of theoretical many-body physics. It has been
ranked along with the band theory of solids and the Landau theory of Fermi
liquids, both single-particle formalisms. The central concept of the BCS
theory is that of fermion \textit{pairings}. In the original model of Cooper
\cite{Coo} they were simply two-electron bound states relative to a full
Fermi sea of the many-electron system. In the BCS theory this original
concept was incorporated into a many-body ground-state variational trial
wavefunction in which all electrons share \textquotedblleft pairing
correlations.\textquotedblright\ The theory, though valid only for
weak-coupling, not only provided a microscopic model for superconductivity,
but it also made many highly specific and quantitative predictions including
explaining the isotope effect, predicting the $T=0$ energy gap $\Delta (0)$\
obeying the universal relation, $2\Delta (0)\simeq 3.53 \ k_{B}T_{c}$ where the
transition temperature $T_{c}$ is the smallest solution of $\Delta
(T_{c})=0, $ and in explaining $T$-dependences of ultrasonic attenuation and
nuclear magnetic resonance relaxation rates \cite{Schrieffer}. Stronger
coupling would not guarantee higher-than-pair clusterings, e.g., tetramers,
etc., since these higher-order charge clusters have not been detected at all
in magnetic-flux quantization measurements in either conventional
superconductors (specifically tin \cite{classical} and lead \cite{classical2}%
) nor in so-called \textquotedblleft high-$T_{c}$\textquotedblright\
compounds such as YBaCuO \cite{cuprates}: the smallest flux unit observed is
$hc/2e$ rather than the originally expected $hc/e$ and it is never \textit{%
smaller} than $hc/2e$.

For many decades the BCS theory, including its extensions into the
strong-coupling regime, appeared to be capable of explaining all of the then
known superconducting elements and compounds. This situation continued while
the highest $T_{c}$ value for any superconductor (SC) was $23$ K, until the
discovery \cite{BM86} in 1986 of the first high-$T_{c}$ cuprate SC %
La$_{2-x}$Ba$_{x}$CuO$_{4}$ having a $T_{c}\simeq 35$ K. The discovery \cite{chu} of
superconductivity at 92 K in YBa$_{2}$Cu$_{3}$O$_{7-\delta }$ was
followed by a search for materials with even higher $T_{c}$s and lead,
 within
just seven years to the highest-$T_{c}$ superconductor known and fully
confirmed to date, the HgBaCaCuO cuprate \cite{Chu93} with a $T_{c}\simeq
164$ K under very high pressure ($\simeq 310,000$ atm).

Almost a quarter century after the discovery of high-temperature SCs in
cuprate materials, it is clear that many important questions still remain to
be answered. As well as the still unresolved problem of the pairing
dynamical mechanism and many-body excitations in the normal state of the
high $T_{c}$ cuprate materials, there are now also many other recently
discovered materials where it is unlikely that BCS theory is applicable, at
least in its original form. These include oxide materials (such as the cubic
bismuthate Ba$_{1-x}$K$_{x}$BiO$_{3}$), borides (such as MgB$_{2}$),
borocarbides (e.g. YNi$_{2}$B$_{2}$C), carbon-based materials (including
fullerides,\ nanotubes, intercalated graphite, and organic conductors), and
new high pressure phases of elements \cite{Buzea05}\ (such as Fe, S and Ca)
and simple binary and ternary compounds. All of these classes of materials
have shown superconductivity above 10 K, including several up to nearly 40
K. Superconductivity at up to 84 K has even been reported in a cubic
ruthenate \cite{Wu97a,Wu97b,blackstead}.

The ongoing debate about the pairing dynamical mechanism in cuprate high $%
T_{c}$ materials has broadly led to two main schools of thought. On the one
hand P.W. Anderson argued from the very beginning \cite{Anderson} that
cuprate materials are in a completely different class from other
superconducting materials, and as such they must have a completely new
pairing mechanism quite different from the BCS theory. In addition to his
original \textquotedblleft resonating-valence-bond\textquotedblright\ (RVB)
model a large range of theories have focussed on superconductivity driven
chiefly by \textit{repulsive} interactions dominated by the\ on-site
Coulomb-repulsion Hubbard $U$. These include gauge theories \cite{lee},
spin-fluctuation theories \cite{bulut,manske,chubukov}, and the
\textquotedblleft Gossammer superconductivity\textquotedblright\ picture of
Laughlin \cite{laughlin}. The discovery of a $d_{x^{2}-y^{2}}$ symmetry
order parameter \cite{annett,tsuei} is generally consistent with pairing
mechanisms deriving from a large positive $U$, and there is some numerical
evidence for a $d_{x^{2}-y^{2}}$ symmetry ground state in the
two-dimensional square lattice Hubbard model \cite{maier}. However, it
remains unclear whether the positive $U$ Hubbard model alone can describe
the hugely complex normal and superconducting state phenomenology of the
cuprate materials \cite{ott} including the characteristic doping
dependences, pseudogaps, marginal Fermi liquid normal state, isotope
effects, and lattice inhomogeneities such as stripes.

On the other hand, many others have taken the view that it is not a
completely new theory that is needed, but rather that the BCS theory should
be extended and/or generalized to describe these new materials. This
approach has the advantage of building upon the foundations of BCS, and
furthermore does not necessarily imply that cuprate superconductivity is in
a completely new class of SCs. Rather, they may be related to other
materials but just in a new parameter regime where the usual approximations
of BCS (even including Eliashberg strong-coupling corrections) may not be
adequate. Some of the many theoretical models which have been examined in
this context include: boson-fermion models \cite{BF5,BF51,BF6,BF61,BF3,BF31},
bipolarons \cite{alexandrov}, the \textquotedblleft pre-formed
pair\textquotedblright\ or BCS-BEC crossover scenario \cite{Eagles69,MRR,MRB2005,levin,PhysicaC07}, non-adiabatic superconductivity \cite{pietronero}, and
generalized Bose-Einstein condensation of Cooper pairs \cite{BF7a,PLA2,CMT02}.

In this work we study electron pairing in a 1D Fermi gas under zero- and finite-range, two-fermion interactions to address some novel and unique properties of CPs in 1D as compared
with the 2D \cite{A-C} and 3D \cite{S-M} cases. In Section 2 we review the Cooper pairing mechanism in a simplified model where two particles near a static Fermi level interact at $T=0$. In Section 3 we obtain \textit{exact} solutions for different types of separable interactions. The properties of the bound pairs are discussed in Section 4 where we derive the dispersion relation for arbitrary values of the center-of-mass momenta (CMM) $\hbar K$. Section 5 presents results for pairs of particles interacting in a more general case where a range parameter is introduced. Section 6 discusses our conclusions.

\section{Cooper pairing}

To define the original Cooper-pair (CP) problem \cite{Coo} consider a system
of $N$ identical fermions in $d$-dimensions interacting through an attractive two-body
potential to study the effects of pairing at nonzero total or center-of-mass
momenta under different types of interaction. At zero temperature, we
assume that the background $(N-2)$-particle system is in the ground state of
an ideal Fermi gas with interactions occurring only in the vicinity of the
Fermi level. The Schr\"{o}dinger equation for two particles in momentum
space is
\begin{equation}
(\mathbf{p}_{1}^{2}/2m+\mathbf{p}_{2}^{2}/2m-E)\Phi (\mathbf{p}_{1},\mathbf{p}_{2})+\sum_{%
\mathbf{p}_{1}^{\prime }\in \mathcal{R}}\sum_{\mathbf{p}_{2}^{\prime }\in
\mathcal{R}}V(\mathbf{p}_{1},\mathbf{p}_{2};\mathbf{p}_{1}^{\prime },\mathbf{%
p}_{2}^{\prime })\Phi (\mathbf{p}_{1}^{\prime },\mathbf{p}_{2}^{\prime })=0
\label{Fullschr}
\end{equation}%
where $\mathcal{R}$ denotes the region of available states above the Fermi
level and $V$ is the two-body interaction. Introducing the CMM $\mathbf{P}$\
and relative momentum $\mathbf{p}$ as%
\begin{equation}
\mathbf{P\equiv }\text{ }\hbar \mathbf{K}=\hbar \mathbf{k}_{1}+
\hbar \mathbf{%
k}_{2}\ \ \ \ \text{and}\ \ \ \ \mathbf{p\equiv }\text{ }\hbar \mathbf{k}=%
\frac{1}{2}\hbar (\mathbf{k}_{1}-\mathbf{k}_{2})  \label{PpKk}
\end{equation}%
the equation of motion for two particles above the Fermi level becomes%
\begin{equation}
{\left[ \hbar ^{2} \mathbf{k}^{2}/m+\hbar ^{2}\mathbf{K}^{2}/4m-E_{K}\right] }\Phi (\mathbf{k,K}%
)+
\sum_{k^{\prime }>k_{F}}V_{\mathbf{k,k}^{^{\prime }}}^{\mathbf{K}}\Phi (%
\mathbf{k}^{\prime },\mathbf{K})=0,  \label{Cooper}
\end{equation}%
where $\Phi (\mathbf{k,K})$ is the two-particle wavefunction. When the
interaction $V_{\mathbf{k,k}^{^{\prime }}}^{\mathbf{K}}\equiv V(\mathbf{K/}2%
\mathbf{+k,}$ $\mathbf{K/}2\mathbf{-k;}$ $\mathbf{K/}2\mathbf{+\mathbf{k}%
^{\prime },}$ $\mathbf{K/}2\mathbf{-k}^{\prime })$ commutes with the CMM
operator associated with $\hbar \mathbf{K}$, $\Phi (\mathbf{k,K})=\varphi (%
\mathbf{k})\Psi (\mathbf{K})$ and the center-of mass wavefunction $\Psi (%
\mathbf{K})$ can be factored out. In 3D $V_{\mathbf{k,k}^{^{\prime }}}^{%
\mathbf{K}}$ can be expanded in partial waves as in the case of the Cooper
model interaction where\ already the $l=0$ contribution leads to a
correlated ground-state with paired particles at the Fermi surface providing
the crucial ingredient in the formulation of the BCS theory \cite{bcs}\ of
superconductivity.

Here we focus on some novel and unique properties of CPs in 1D, as compared
with the 2D \cite{A-C} and 3D \cite{S-M} cases. We start from a nonlocal,
separable interaction given by%
\begin{equation}
V_{\mathbf{k},\mathbf{k}^{\prime }}^{\mathbf{K}}=-V_{0}g(k)g(k^{\prime })
\label{sepv}
\end{equation}%
with $V_{0}>0.$ This includes all cases of physical interest such as the
Cooper \cite{Coo}, the BCS \cite{bcs}, the zero-range or contact \cite{A-C,S-M,M-C,A-D} model interactions, as well as the finite-range model
interactions as introduced by Nozi\`{e}res and Schmitt-Rink \cite{S-R}\ that
have been used in the description of superconductors, superfluids,
Bose-Einstein condensates, as well as the BCS-BEC crossover \cite{Eagles69,MRR,MRB2005,levin,PhysicaC07} picture. The separable form (\ref{sepv}) assumes that
such an expression is valid in each partial-wave channel and may vary for
angular momentum states different from zero in 2D and 3D. In all these
cases, (\ref{Cooper}) has an analytical solution given by%
\begin{equation}
\varphi (k)=\frac{V_{0}g(k)A}{\hbar ^{2}k^{2}/m+\hbar ^{2}K^{2}/4m-E_{K}}
\label{phi}
\end{equation}%
where%
\begin{equation*}
A\equiv \sum_{k^{\prime }>k_{F}}g(k^{\prime })\varphi (k^{\prime })
\end{equation*}%
is a constant. The combination of these two equations provides a consistency
condition that is equivalent to an eigenvalue equation for ${E}_{K}$, namely%
\begin{equation}
1=V_{0}\sum_{k>k_{F}}\frac{g^{2}(k)}{\hbar ^{2}k^{2}/m+\hbar
^{2}K^{2}/4m-E_{K}}.  \label{Sch}
\end{equation}%
Defining $-\Delta _{K}\equiv E_{K}-2E_{F}$ as the pair energy with respect
to twice the Fermi energy $E_{F}\equiv \hbar ^{2}k_{F}^{2}/m$,
the continuous limit for large $N$ and large volume is%
\begin{equation}
1=V_{0}\left( {2\pi }\right) ^{-d}L^{d}\int {\frac{d^{d}kg^{2}(k)}{{\hbar
^{2}(k^{2}-k_{F}^{2})/m+\Delta _{K}+\hbar ^{2}K^{2}/4m}}}  \label{eigen1}
\end{equation}%
where $L^{d}$ is the \textquotedblleft volume" of the $d$ dimensional system.

When the interaction occurs only in the vicinity of the Fermi energy $%
E_{F}\equiv {\hbar ^{2}k_{F}^{2}/2m}$, and since the integrand in (\ref%
{eigen1}) is peaked at $k\sim k_{F}$, the integral can be reexpressed in
terms of the density of states (DOS) for one spin at the Fermi level, namely%
\begin{equation}
\varrho _{d}(E_{F})=m\left( \frac{L}{2\pi \hbar }\right) ^{d}c_{d}\left(
2mE_{F}\right) ^{d/2-1},  \label{dos}
\end{equation}%
where $c_{d}=4\pi ,$ $2\pi ,2$ for $d=3,2,1,$
respectively.

In 1D%
\begin{equation}
\varrho _{1}(E_{F})=\frac{mL}{\pi \hbar ^{2}k_{F}}.  \label{dos1}
\end{equation}%
If $g(k)$ is constant around the Fermi level the usual Cooper binding-energy
result can be obtained for any CMM wavenumber $K\geqslant 0$. However, this
approximation is unnecessary since one can analytically integrate (\ref%
{eigen1}) for most interaction models of physical interest. For brevity, one
can express all energies in terms of the Fermi energy and all wavenumbers in
units of the Fermi wavenumber $k_{F}$, namely through dimensionless $\tilde{K%
}=K/k_{F},$ $\tilde{k}=k/k_{F}$ and $\tilde{\Delta}_{K}=\tilde{\Delta}%
_{K}/E_{F}$. Then (\ref{eigen1}) becomes%
\begin{equation}
1=\frac{\lambda }{2}\int \frac{d\tilde{k}g^{2}(\tilde{k})}{\tilde{k}^{2}+%
\tilde{\Delta}_{K}{/2+}\tilde{K}^{2}/4-1}  \label{eigen2}
\end{equation}%
where the dimensionless coupling constant is
\begin{equation}
\lambda \equiv \varrho _{1}(E_{F})V_{0}=\frac{V_{0}mL}{\pi \hbar ^{2}k_{F}}.
\label{dimlesscoupl}
\end{equation}

Condition (\ref{eigen2}) is the effective characteristic equation for the bound-pair energy $\Delta_{K}$ which can be solved exactly for any form factor $g(k)$ of physical interest.
\section{Cooper pairing in one dimension}
\begin{figure}
\includegraphics[angle=-90, scale=0.5]{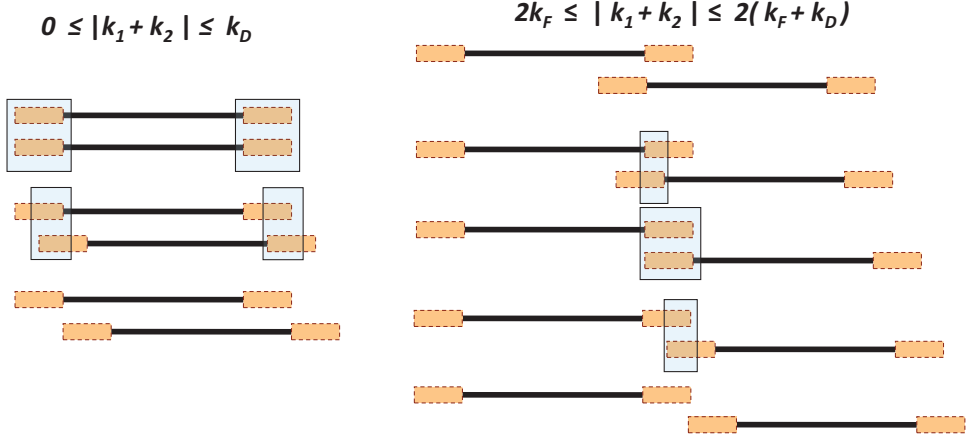}
\caption{\textquotedblleft Dumbbell\textquotedblright\ picture of BCS or Cooper interaction regions (shaded boxes, not in scale) in momentum space for pairs of
fermions. Dumbbells are laterally separated by the total CMM $K$ of a CP.
Uppermost part in the left block represents $K=0$ where the available phase-space for
interactions is maximum. In contrast with higher-dimensional systems, in 1D there is an interval where the interaction phase space vanishes as $K$ increases from $K=k_{D}$ to $K=2k_{F}$ (right
block of diagrams). }
\label{fig:listones}
\end{figure}

Fermi systems in 1D reveal a novel multiphase pairing mechanism that is
absent in higher-dimensional systems. When pair interactions act only in the
vicinity of the Fermi energy as in the case of the Cooper \cite{Coo} and the
BCS \cite{bcs} model interactions, the available phase space has at least
two discontinuous intervals for the total or CMM $\hbar K,$ see Fig. \ref%
{fig:listones}. When $K\ll 2k_{F},$\ particles at both ends of the
\textquotedblleft dumbbell\textquotedblright\ feel the attractive
interactions because their energies are close to the Fermi energy $E_{F}.$\
As $K$\ increases there is a region where particles no longer interact due
to their large momenta, until the condition $k_{F}<K/2<k_{F}+k_{D}$\ is
satisfied, where $k_{D}\equiv k_{F}\sqrt{1+\hbar \omega _{D}/E_{F}}-k_{F}$\
as shown in the lower part of Fig. \ref{fig:listones}. The cutoff
wavenumber $k_{D}$\ is related to the maximum ionic-lattice vibrational (or
Debye) frequency $\omega _{D}$\ through $\hbar
^{2}(k_{F}+k_{D})^{2}/2m\equiv E_{F}+\hbar \omega _{D}.$ Clearly, for a
contact interaction $\hbar \omega _{D}/E_{F}\rightarrow \infty $ so that $%
k_{D}\rightarrow \infty .$

The form factors $g(k)$ of the separable pairing interaction (\ref{sepv})
are given by%
\begin{equation}
g(k)=\left\{
\begin{array}{ll}
\theta (\epsilon _{1,2}-E_{F}) & \text{ \ \ contact interaction} \\
\theta (\epsilon _{1,2}-E_{F})\theta (E_{F}+\hbar \omega _{D}-\epsilon
_{1,2}) & \text{ \ \ Cooper interaction} \\
\theta (\epsilon _{1,2}-[E_{F}-\hbar \omega _{D}])\theta ([E_{F}+\hbar
\omega _{D}]-\epsilon _{1,2}) & \text{ \ \ BCS interaction}%
\end{array}%
\right.   \label{gofk3}
\end{equation}%
where
\begin{equation*}
\theta (x)\equiv \left\{
\begin{array}{l}
1\text{ for }x>0 \\
0\text{ for }x<0%
\end{array}%
\right.
\end{equation*}%
and $\epsilon _{1,2}$ are the energies of particle $1$ and of particle $2$
given by%
\begin{equation*}
\epsilon _{1,2}\equiv \hbar ^{2}k_{1,2}^{2}/2m\equiv \hbar ^{2}\left( K/2\pm
k\right) ^{2}/2m.
\end{equation*}%
In terms of the wavenumber, the form factors in (\ref{gofk3}) are%
\begin{equation}
g(\tilde{k})=\left\{
\begin{array}{ll}
\theta (\left\vert \tilde{k}\right\vert -[1+\tilde{K}/2]) & \text{ \ \
contact interaction} \\
\theta (\left\vert \tilde{k}\right\vert -[1+\tilde{K}/2])\theta (\sqrt{%
1+\hbar \omega _{D}/E_{F}}-\tilde{K}/2-\left\vert \tilde{k}\right\vert ) &
\text{ \ \ Cooper interaction} \\
\theta (\left\vert \tilde{k}\right\vert -[\sqrt{1-\hbar \omega _{D}/E_{F}}-%
\tilde{K}/2])\theta ([\sqrt{1+\hbar \omega _{D}/E_{F}}-\tilde{K}%
/2]-\left\vert \tilde{k}\right\vert ) & \text{ \ \ BCS interaction}%
\end{array}%
\right.   \label{gofk4}
\end{equation}%
where, for definiteness, we have assumed $\tilde{K}>0$ since the $\tilde{K}<0
$ sector provides the same results. In what follows we occasionally drop the
tildes.%

The BCS interaction poses a problem that was first mentioned by Schrieffer (%
Ref. [3], p. 168) related to the singularities in (\ref{eigen2})
since the integration intervals defined in (\ref{gofk4}) may produce zeros
in the denominator. However, for contact and Cooper interactions (\ref%
{eigen2}) can be solved exactly yielding, for $K\ll 1,$%

\begin{equation}
\exp (-2\alpha _{K}/\lambda )=
\left\{
\begin{array}{ll}
\frac{(1+K/2-\alpha _{K})}{(1+K/2+\alpha _{K})} & \text{contact interaction} \\
\frac{(1+k_{D}-K/2+\alpha _{K})(1+K/2-\alpha _{K})}{(1+k_{D}-K/2-\alpha
_{K})(1+K/2+\alpha _{K})} & \text{Cooper interaction}%
\end{array}%
\right.  \label{res}
\end{equation}%

where $k_{D}$ was defined above and
\begin{equation}
\alpha _{K}^{2}\equiv 1-K^{2}/4-\Delta _{K}/2.  \label{alphaK}
\end{equation}%

For $K$ sufficiently large $\alpha _{K}^{2}$ becomes negative and the
existence of stable solutions for ${\Delta _{K}}$ depends on the magnitude
of $\lambda .$ For $\lambda \ll 1$ one has, if $\beta
_{K}^{2}\equiv -\alpha _{K}^{2}$,%
\begin{equation}
\frac{\beta _{K}}{\lambda }=\left\{
\begin{array}{ll}
\pi /2-\tan ^{-1}\left[ (1+K/2)/\beta _{K}\right] & \text{ \ \ contact
interaction} \\
\tan ^{-1}\left[ (1+k_{D}-K/2)/\beta _{K}\right] -\tan ^{-1}\left[
(1+K/2)/\beta _{K}\right] & \text{ \ \ Cooper interaction}%
\end{array}%
\right.  \label{res2}
\end{equation}%
Since (\ref{res}) are transcendental in the CP energy ${\Delta _{K}}$ (%
\textit{not }to be confused with the BCS energy gap \cite{bcs}), it is
customary to consider the case where the largest number of particles are
interacting, namely when $K=0$ as shown in Fig. \ref{fig:listones}. In this
approximation, one introduces the DOS for one spin $\varrho _{1}(\epsilon )$%
\ in (\ref{eigen2}) and assumes that it is constant around the Fermi
surface. This is strictly true in 2D and otherwise a good approximation in
any D provided that $\hbar \omega _{D}/E_{F}\ll 1$. Then, the binding energy
$\Delta _{0}$ for $K=0$ CPs turns out to be given by the familiar limit for
the Cooper interaction \cite{Coo}%
\begin{equation}
\Delta _{0} \ \xrightarrow[\lambda \rightarrow 0]{} \  2 \hbar \omega
_{D}\exp (-2/\lambda )
\label{gap}
\end{equation}%
where $\lambda \equiv \varrho_1 (E_{F})V_{0}$ is the usual dimensionless
coupling parameter. Note that the CP $\Delta _{0}$ vanishes like $\exp
(-2/\lambda )$ unlike the BCS gap \cite{bcs} which vanishes like $\exp
(-1/\lambda )$ $.$

For a contact interaction of the form $-v_{0}\delta (x),$ where $x$\ is the
separation between the two fermions, the strength $v_{0}$ has dimensions of
energy $\times $ length and the pair binding energy can again be calculated
using the same approximation, namely taking the DOS at the Fermi level.
After some manipulation the small-coupling limit of $\Delta _{0}$ is (see
Ref. \cite{M-C}\ and esp. Ref. \cite{A-D} Eq. 17)%
\begin{equation}
\Delta _{0}\ \xrightarrow[\lambda \rightarrow 0]{} \ 8E_{F}\exp
(-2/\lambda )  \label{gapcontact}
\end{equation}%
where in this case the dimensionless coupling parameter is given by $\lambda
\equiv \varrho_1 (E_{F})v_{0}/L$. This same result also follows from the
Cooper interaction result in (\ref{res2}) on taking the limit $%
k_{D}\rightarrow \infty $.

\section{CP dispersion relation}

\begin{figure}[tbh]
\begin{center}
\centerline{\epsfig{file=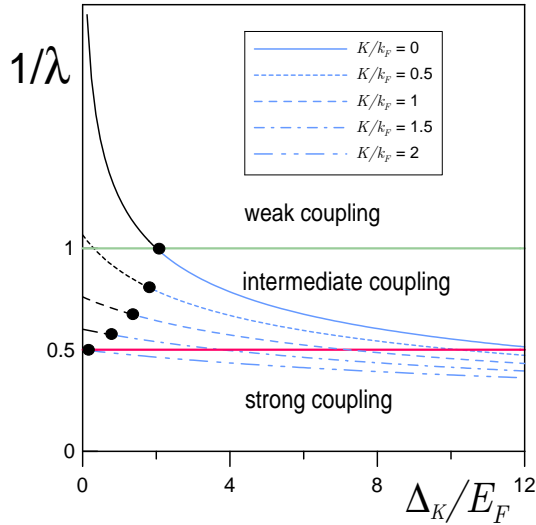,height=2.80in,width=2.80in}}
\end{center}
\par
\vspace{-1.0cm}
\caption{Graphical determination of CP energies $\Delta _{K}$ for different
strengths $\protect\lambda $ in (\protect\ref{contsols}). Horizontal lines
are lhs and curves are rhs of (\protect\ref{contsols}) for five different
values of $K$. The solution $\Delta _{K}$ is given by the abscissa where the
corresponding $K$-curve crosses (dots) the $1/\protect\lambda $ line. Note
that there are some intervals of $\protect\lambda $ and of $K$ without
solutions. Black dots on curves associated with a fixed $K$ define points at
which the analytic forms on rhs of (\protect\ref{contsols}) change from one
to another.}
\label{LAMBDAVSDEL}
\end{figure}

We analyze the case of a contact interaction where the solutions are given
in (\ref{res})-(\ref{res2}).\texttt{\ }As mentioned, the analytical
solutions depend on the strength of the dimensionless coupling $\lambda $\
and on the magnitude of $K.$\ When $K>2$\ one must take into account an
additional restriction due to the integration regions defined in (\ref{gofk4}%
) since for stronger coupling one would expect larger values for $\Delta
_{K}.$\ Therefore, for a particular $\lambda $\ we define the value $%
K_{c}(\lambda )$\ for which $\alpha _{K}\mid _{K=K_{c}}=0$ implying from (%
\ref{alphaK})\ that $K_{c}^{2}/4\equiv 1-\Delta _{K_{c}}/2$\ where $\Delta
_{K_{c}}$\ is the CP energy for $K=K_{c}.$\ The transcendental equations for
the CP energy $-\Delta _{K}$ are then, again if $\beta _{K}^{2}\equiv
-\alpha _{K}^{2}$,
\begin{equation}
\frac{1}{\lambda }=\left\{
\begin{array}{lll}
-\frac{1}{2\alpha _{K}}\ln [(1+K/2-\alpha _{K})/(1+K/2+\alpha _{K})]\,\  &
\text{for }K<2,\text{ \ \ }0<\lambda <2 & \text{\quad a) }\vspace{0.2cm} \\
\frac{1}{\beta _{K}}\left[ \pi /2-\tan ^{-1}([1+K/2]/\beta _{K})\right]
\qquad & \text{for }K<2,\text{ \ \ }2<\lambda <\infty & \text{\quad b) }%
\vspace{0.2cm} \\
-\frac{1}{2\alpha _{K}}\ln [(1+K/2-\alpha _{K})/(1+K/2+\alpha _{K})] & \text{%
for }K_{c}<K<2,\text{ \ \ }1<\lambda <2 & \text{\quad c) }\vspace{0.2cm} \\
\frac{1}{\beta _{K}}(\pi /2-\tan ^{-1}[(1+K/2)/\beta _{K}])\qquad \qquad &
\text{for }0<K<K_{c},\text{ \ \ }1<\lambda <2 & \text{\quad d) }\vspace{0.2cm%
} \\
\frac{1}{\beta _{K}}(\pi /2-\tan ^{-1}[(K/2+1)/\beta _{K}]+\tan
^{-1}[(K/2-1)/\beta _{K}]) & \text{for }K>2,\text{ }\lambda >0. & \text{%
\quad e)}%
\end{array}%
\right.  \label{contsols}
\end{equation}%
The existence of analytically different solutions\ follows from the
graphical construct shown in Fig. \ref{LAMBDAVSDEL} where we plot the rhs
of (\ref{contsols}) and $1/\lambda $ as a function of the possible values of
the CP energy $\Delta _{K}$\ for different values of $K.$ For convenience,
we label as type-I those solutions that satisfy (\ref{contsols}a) and (\ref%
{contsols}c); the other solutions are labeled type-II\textit{. }Solutions $%
\Delta _{K}$ only exist at those points where the $K$-curves cross the $%
1/\lambda $ lines. From this construct three regions can be distinguished:

a) \textbf{Weak coupling} $0<\lambda <1$ and $(\Delta _{K}/2+K^{2}/4-1)<0.$
In this region, there will always be a type-I solution for small values of $%
K.$

b) \textbf{Intermediate coupling }$1<\lambda <2$ where $\alpha _{K}^{2}$
changes sign and therefore both types of solution are present.

c) \textbf{Strong coupling} $\lambda >2$ and $(\Delta _{K}/2+K^{2}/4-1)>0.$
In this region, only type-II solutions are possible.

For $K=0$ (\ref{contsols}a) reduces to%
\begin{equation}
\exp (-\frac{2}{\lambda }\sqrt{1-{\Delta _{0}/}2})=\frac{{1-\sqrt{1-{\Delta
_{0}/}2}}}{{1+\sqrt{1-{\Delta _{0}/}2}}}.  \label{eqeB}
\end{equation}
\begin{figure}[tbh]
\begin{center}
\centerline{\epsfig{file=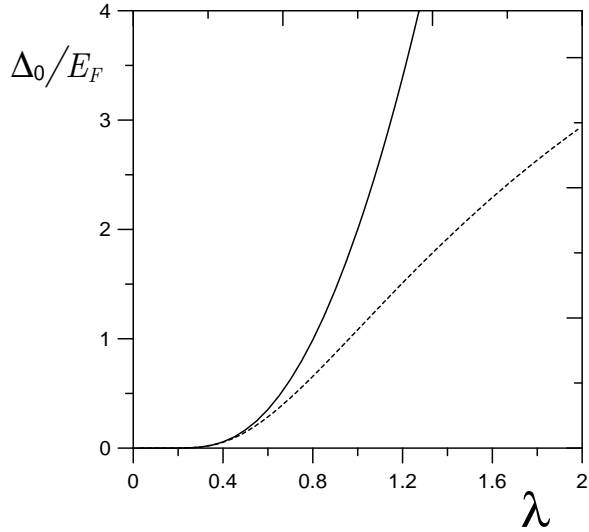,height=2.80in,width=3.1in}}
\end{center}
\par
\vspace{-0.50cm}
\caption{Exact $K=0$ CP energy $\Delta _{0}$\ for the contact interaction as
function of coupling $\protect\lambda $ (full curve) as obtained from (%
\protect\ref{eqeB}) compared to the weak-coupling approximation (\protect\ref%
{gapcontact}) (dashed curve).}
\label{DeltaCeroII}
\end{figure}
\begin{figure}[htb]
\begin{center}
\centerline{\epsfig{file=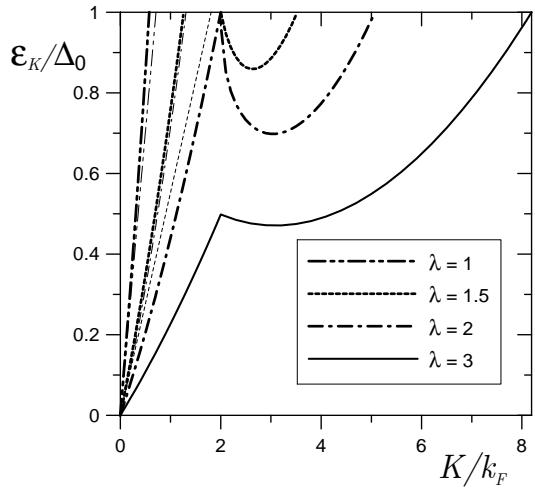,height=2.60in,width=2.80in}}
\end{center}
\par
\vspace{-0.50cm}
\caption{Pair excitation energy $\mathcal{E}_{K}\equiv \Delta _{0}-\Delta
_{K}$ in units of $\Delta _{0}$ for a contact interaction for different
values of $\protect\lambda $ (thick curves). Thin curves refer to pure
linear while thin dot-dashed curve for $\protect\lambda =1.5$ is the linear
plus quadratic behavior. For weak-coupling ($\protect\lambda \leq 1$), only
the linear phonon-like term is present with the pair breaking up whenever $%
\mathcal{E}_{K}\geq 1$. For larger coupling, e.g., $\protect\lambda \geq 1.5$%
, a roton-like branch appears when $\tilde{K}\geq 2$.}
\label{EnVSKDELII}
\end{figure}
As a consistency check we determine the CP energy $\Delta _{K}$ when the
Fermi sea vanishes, i.e., $E_{F}\rightarrow 0$ and $k_{F}\rightarrow 0$ or
when the CP is \textit{in vacuo}. We note that only (\ref{contsols}e) is
meaningful in this limit since $\tilde{K}<2$ implies $K<2k_{F}$ in
wavenumber units so that $K\leq 0$ when $k_{F}\rightarrow 0$ which
contradicts the assumption made just below (\ref{gofk4})\ that $K>0.${\Huge %
\ }Introducing in (\ref{contsols}e) the value of $\lambda $\ defined in (\ref%
{dimlesscoupl}) in terms of $k_{F}$\ and reverting to explicit energy and
wavenumber units one obtains%
\begin{equation}
\frac{\beta _{K}}{\lambda }=\pi /2-\tan ^{-1}[(K/2k_{F}+1)/\beta _{K}]+
\tan^{-1}[(K/2k_{F}-1)/\beta _{K}] \nonumber
\end{equation}%
or%
\begin{eqnarray*}
\frac{\sqrt{\Delta _{K}m/\hbar ^{2}+K^{2}/4-k_{F}^{2}}}{mV_{0}/\pi \hbar ^{2}}=
\pi /2-\tan ^{-1}\left[ (K/2+k_{F})/\sqrt{\Delta _{K}m/\hbar
^{2}+K^{2}/4-k_{F}^{2}}\right] \\
+\tan ^{-1}\left[ (K/2-k_{F})/\sqrt{\Delta _{K}m/\hbar
^{2}+K^{2}/4-k_{F}^{2}}\right] .
\end{eqnarray*}%
When $k_{F}\rightarrow 0$\ this leads to%
\begin{equation*}
-\Delta _{K}=-\frac{mV_{0}^{2}}{4\hbar ^{2}}+\frac{\hbar ^{2}K^{2}}{4m}
\end{equation*}%
which is the expected actual energy of a composite object of mass $2m$,
self-bound via a 1D delta potential of \textit{arbitrary} strength $V_{0}$,
with its single-bound-state binding energy\cite{gasior,Ll01} $mV_{0}^{2}/4\hbar ^{2}$  and moving freely \textit{in vacuo}.

When $K\ll 1$ we may assume the series expansion $\Delta _{K}\simeq \Delta
_{0}+\Delta _{1}K+\Delta _{2}{K}^{2}+\cdots $ in (\ref{eqeB}). For weak
coupling, $\lambda \rightarrow 0$\ implies that ${\Delta _{0}\longrightarrow
0}$\ so that $\sqrt{1-{\Delta _{0}/}2}\simeq 1-{\Delta _{0}/}4$ ${+}$ ${%
\cdots }$ which leads to (\ref{gapcontact}) as expected. The same result is
obtained using the DOS approximation given in (\ref{gap}) above. In Fig. \ref%
{DeltaCeroII} we show the exact result for ${\Delta }_{0}$ compared to (\ref%
{gapcontact}) for weak coupling. The coefficient of the linear term $\Delta
_{1}$ is of special interest. It can be obtained explicitly if we assume
weak coupling and using the fact that $\Delta _{0}$ can be neglected in the
exponentials, that is, by assuming that $\exp (-2\alpha _{K}/\lambda )$%
{\Huge \ }$\simeq ${\Huge \ }$\exp (-2/\lambda )$. The result is%
\begin{equation}
\Delta _{1}\simeq -2\tanh ^{2}\left( {1/}\lambda \right) .  \label{delta1}
\end{equation}%
A similar procedure can be used to obtain the second-order coefficient%
\begin{equation}
\Delta _{2}\simeq -\tanh ^{4}\left( {1/}\lambda \right) .  \label{delta2}
\end{equation}%
Therefore, the dispersion relation for ${\Delta _{K}}$ up to second order in
$K$ but for weak coupling $\lambda $\ in explicit energy and wavenumber
units is%
\begin{eqnarray}
\Delta _{K} &\simeq &{\Delta }_{0}-\tanh ^{2}\left( {1/}\lambda \right)
\hbar v_{F}K-{{\tanh }^{4}}\left( {1/}\lambda \right) \frac{{\hbar ^{2}K^{2}}%
}{{2m}}+\cdots  \notag \\
&& \hspace{-0.50cm} \xrightarrow[\lambda \rightarrow 0]{} \ 8E_{F}\exp (-2/\lambda
)-\hbar v_{F}K-\frac{{\hbar ^{2}K^{2}}}{{2m}}+\cdots  \label{expansiondeltak}
\end{eqnarray}%
where (\ref{gapcontact}) was used. The negative signs in the first- and
second-order terms for a given coupling $\lambda $ implies that the pair
will break-up as $K$ increases beyond a certain value for which $\Delta
_{K}=0.$ However, for $K\geq 2k_{F}$ and for sufficiently large $\lambda $,
a pair can become bound again as shown in Fig. \ref{EnVSKDELII} where we
plot the \textit{gapped} excitation energy $\mathcal{E}_{K}\equiv \Delta
_{0}-\Delta _{K}$ as a function of $K$ for different couplings. The special
case of $\lambda =1.5$ illustrates this behavior. As $K$ increases from
zero, the excitation energy is essentially linear up to $K\simeq 1.2k_{F}$
when the pair breaks up but for $K\geq 2k_{F}$ the pair comes back into
existence with an excitation energy that is close to quadratic in the pair
wavenumber $K$.
\section{Finite-range interactions}
In contrast with the cases mentioned above, here we consider a more general
interaction between fermions where a range parameter is introduced albeit
the interaction form is still separable. In 3D it is customary to introduce
a screened interaction of the Yukawa form $\exp (-r/r_{0})/r$ which in
momentum space is ${\Large \varpropto }$ $[(q^{2}+(1/r_{0})^{2}]^{-2}$,
where $q$ is the momentum transfer wavenumber. Based on this criterion and
following previous calculations (cf. Ref. \cite{S-R} 
esp. Eq. 10) we write
the form factor $g(k)$ in (\ref{sepv}) as
\begin{equation}
g(k)=\frac{\theta (\left\vert k\right\vert -[1+K/2])}{\sqrt{k^{2}+k_{0}^{2}}}
\label{gfinite}
\end{equation}%
where all wavenumbers are again in units of $k_{F}.$ The characteristic
equation for the energy (\ref{eigen2}) becomes%
\begin{equation}
\frac{1}{\lambda }=\int_{1+K/2}^{\infty }\frac{dk}{\left( k^{2}+\Delta
_{K}/2+K^{2}/4-1\right) \left( k^{2}+k_{0}^{2}\right) }.  \label{eqrange}
\end{equation}%
This equation can be solved exactly and leads to the transcendental
equations for the CP energy%
\begin{subequations}
\begin{align}
\frac{(k_{0}^{2}+\alpha _{K}^{2})}{\lambda }& =-\frac{1}{2\alpha _{K}}\ln %
\left[ \frac{1+K/2-\alpha _{K}}{1+K/2+\alpha _{K}}\right] -\frac{\pi }{2k_{0}%
}+\frac{1}{k_{0}}\tan ^{-1}\left[ \frac{1+K/2}{k_{0}}\right] & K/2& \ll 1,%
\text{ \ }\alpha _{K}^{2}>0\text{ }  \label{resfinite1} \\
\frac{k_{0}^{2}-\beta _{K}^{2}}{\lambda }& =\frac{1}{\beta _{K}}\left( \frac{%
\pi }{2}-\tan ^{-1}\left[ \frac{1+K/2}{\beta _{K}}\right] \right) -\frac{1}{%
k_{0}}\left( \frac{\pi }{2}-\tan ^{-1}\left[ \frac{1+K/2}{k_{0}}\right]
\right) & K/2& <1,\text{ \ }\alpha _{K}^{2}<0  \label{resfinite2} \\
\frac{k_{0}^{2}-\beta _{K}^{2}}{\lambda }& =\frac{1}{\beta _{K}}\left( \frac{%
\pi }{2}+\tan ^{-1}\left[ \frac{K/2-1}{\beta _{K}}\right] -\tan ^{-1}\left[
\frac{1+K/2}{\beta _{K}}\right] \right) & &  \notag \\
& \hspace{1.8cm}-\frac{1}{k_{0}}\left( \frac{\pi }{2}+\tan ^{-1}\left[ \frac{%
K/2-1}{k_{0}}\right] -\tan ^{-1}\left[ \frac{1+K/2}{\beta _{K}}\right]
\right) & K/2& >1,\text{ \ }\alpha _{K}^{2}<0  \label{resfinite3}
\end{align}%
\end{subequations}
%
%
%
%
%
%
%
%
%
%
%
%

\begin{figure}[tbh]
\begin{center}
\centerline{\epsfig{file=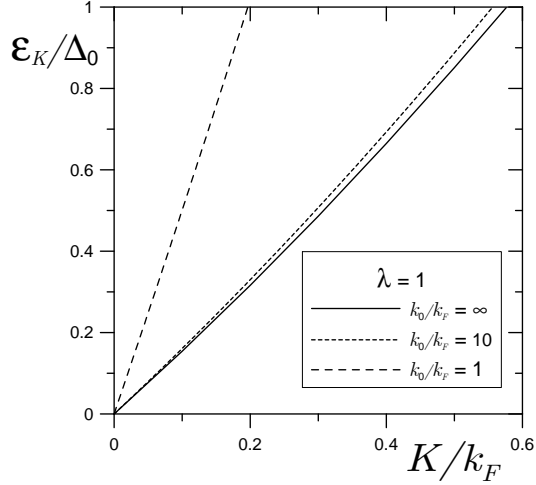,height=2.6in,width=2.8in}}
\end{center}
\par
\vspace{-1.0cm}
\caption{Excitation energy $\mathcal{E}_{K}$ in units of $\Delta _{0}$\ of a
CP for moderate coupling $\protect\lambda =1$ and for different interaction
ranges gauged by the inverse range parameter $k_{0}$. The dispersion
relation is essentially linear over the entire range of $K$ below the
pair-breaking limit $\mathcal{E}_{K}/\Delta _{0}=1.$ Infinite $k_{0}$ is the
contact-interaction limit.}
\label{EnVSKSRL1}
\end{figure}
\begin{figure}[tbh]
\begin{center}
\centerline{\epsfig{file=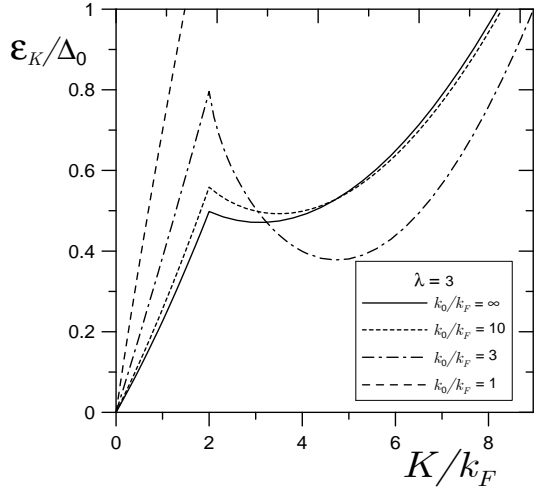,height=2.6in,width=2.8in}}
\end{center}
\par
\vspace{-1.0cm}
\caption{Excitation energy $\mathcal{E}_{K}$ in units of $\Delta _{0}$ of a
CP for strong coupling, $\protect\lambda =3,$ for different values of
inverse range parameter $k_{0}$. For $k_{0}\geq 3$ the two distinct
dispersion branches are possible. Infinite $k_{0}$ refers to the
contact-interaction limit.}
\label{EnVSKSRL3}
\end{figure}
Again, there are three regions for the existence of stable pairs. In Fig. %
\ref{EnVSKSRL1} we plot the pair energy as a function of $K$ for weak
coupling and for different values of the range parameter $1/k_{0}$. Stronger
coupling spectra are shown in Fig. \ref{EnVSKSRL3} where we plot the binding
energies for $\lambda =3$ where both, the linear and roton-like modes are
present \cite{Donn}. The sharp cusp separating phonon-like and roton-like modes at
precisely $K=2k_{F}$ is a unique characteristic of the 1D system. It appears
to be a precursor of the smooth \textquotedblleft
maxon-like\textquotedblright\ hump in 2D (Ref. \cite{A-C}, Figs. 1 and 3) and
less pronounced in 3D (Ref. \cite{S-M}, Fig. 2). 

\section{Conclusions}

Notwithstanding the obvious simplicity of a model consisting of a
many-fermion system where particles interact only in the vicinity of the
Fermi level through two-body, attractive, separable interactions in the
background of an ideal 1D Fermi gas, it reveals novel, unique properties,
particularly those related to the energy of Cooper pairs moving with \textit{%
nonzero} center-of-mass momentum (CMM) $K$. The fact that one can calculate
exact expressions for the pair energy with different separable interactions
allows us to construct the collective excitation spectrum of Cooper pairs
for any coupling $\lambda $\ and any value of $K$. For $K<k_{F}$, the
excitation energy has a linear term in $K$. As the CMM wavenumber $K$
increases the pair eventually breaks up. However, for sufficiently strong
coupling, there is an additional collective mode for $K\geq 2k_{F}$ with a
roton-like dispersion. For a contact interaction, the two modes are
disconnected if the coupling is weak but for stronger coupling the
excitation spectrum exhibits both modes. Introducing more realistic
interactions that include screening effects shows a similar behavior.

The sharp cusp separating phonon-like and roton-like modes at precisely $%
K=2k_{F}$ is a unique characteristic of the 1D system. It appears to be a
precursor of the smooth \textquotedblleft maxon-like\textquotedblright\ hump
in 2D (Ref. \cite{A-C}, Figs. 1 and 3) and less pronounced in 3D (Ref. \cite{S-M}, Fig. 2). The smoothness could be due to angular integrations in 2D and
3D washing out the beaked transition between both modes found here in 1D.

\textbf{Acknowledgments} We acknowledge partial support from grants PAPIIT
IN114708, IN106908 and CONACyT 104917.

\end{document}